\documentclass[sn-mathphys,Numbered]{sn-jnl}

\usepackage{graphicx}%
\usepackage{multirow}%
\usepackage{amsmath,amssymb,amsfonts}%
\usepackage{amsthm}%
\usepackage{mathrsfs}%
\usepackage[title]{appendix}%
\usepackage{xcolor}%
\usepackage{textcomp}%
\usepackage{manyfoot}%
\usepackage{booktabs}%
\usepackage{algorithm}%
\usepackage{algorithmicx}%
\usepackage{algpseudocode}%
\usepackage{listings}%
\usepackage{wrapfig}

\begin{document}

\title[Article Title]{The Data Acquisition System for Phase-III of the BeEST Experiment}

\author*[1,2]{\fnm{C.}~\sur{Bray}}\email{cbray@mines.edu\vspace*{-1.5em}}

\author[1,2]{\fnm{S.}~\sur{Fretwell}}

\author[2]{\fnm{I.}~\sur{Kim}}

\author[3]{\fnm{W.}~K.~\sur{Warburton}}

\author[4]{\fnm{F.}~\sur{Ponce}}

\author[1,5]{\fnm{K.}~G.~\sur{Leach}}

\author[2]{\fnm{S.}~\sur{Friedrich}}

\author[6]{\fnm{R.}~\sur{Abells}}

\author[7]{\fnm{P.}~\sur{Amaro}}

\author[8]{\fnm{A.}~\sur{Andoche}}

\author[9]{\fnm{R.}~\sur{Cantor}}

\author[1]{\fnm{D.}~\sur{Diercks}}

\author[7]{\fnm{M.}~\sur{Guerra}}

\author[9]{\fnm{A.}~\sur{Hall}}

\author[1]{\fnm{C.}~\sur{Harris}}

\author[3]{\fnm{J.}~\sur{Harris}}

\author[10]{\fnm{L.}~\sur{Hayen}}

\author[8]{\fnm{P.~A.}~\sur{Hervieux}}

\author[2]{\fnm{G. B.}~\sur{Kim}}

\author[6]{\fnm{A.}~\sur{Lennarz}}

\author[2]{\fnm{V.}~\sur{Lordi}}

\author[7]{\fnm{J.}~\sur{Machado}}

\author[6]{\fnm{P.}~\sur{Machule}}

\author[1]{\fnm{A.}~\sur{Marino}}

\author[6]{\fnm{D.}~\sur{McKeen}}

\author[11]{\fnm{X.}~\sur{Mougeot}}

\author[6]{\fnm{C.}~\sur{Ruiz}}

\author[2]{\fnm{A.}~\sur{Samanta}}

\author[7]{\fnm{J.}~P.~\sur{Santos}}

\author[1]{\fnm{C.}~\sur{Stone-Whitehead}}

\affil[1]{\orgname{Colorado School of Mines}, \orgaddress{\city{Golden}, \state{Colorado}, \country{USA}}}

\affil[2]{\orgname{Lawrence Livermore National Laboratory}, \orgaddress{\city{Livermore}, \state{California}, \country{USA}}}

\affil[3]{\orgname{XIA LLC}, \orgaddress{\city{Oakland}, \state{California}, \country{USA}}} 

\affil[4]{\orgname{Pacific Northwest National Laboratory}, \orgaddress{\city{Richland}, \state{Washington}, \country{USA}}}

\affil[5]{\orgdiv{Facility for Rare Isotope Beams}, \orgaddress{\city{East Lansing}, \state{Michigan}, \country{USA}}}

\affil[6]{\orgname{TRIUMF}, \orgaddress{\city{Vancouver}, \state{BC}, \country{Canada}}}

\affil[7]{\orgname{NOVA School of Science and Technology}, \orgaddress{\city{Lisbon}, \country{Portugal}}}

\affil[8]{\orgname{Université de Strasbourg}, \orgaddress{\city{Strasbourg}, \country{France}}}

\affil[9]{\orgname{Star Cryoelectronics}, \orgaddress{\city{Santa Fe}, \state{New Mexico}, \country{USA}}}

\affil[10]{Laboratoire de Physique Corpusculaire, CNRS/IN2P3, \orgaddress{\city{Caen}, \country{France}}}

\affil[11]{\orgname{CEA Saclay}, \orgaddress{\city{Paris}, \country{France}}\vspace*{-1.5em}}

\abstract{The BeEST experiment is a precision laboratory search for physics beyond the standard model that measures the electron capture decay of  $^7$Be implanted into superconducting tunnel junction (STJ) detectors. For Phase-III of the experiment, we constructed a continuously sampling data acquisition system to extract pulse shape and timing information from 16 STJ pixels offline. Four additional pixels are read out with a fast list-mode digitizer, and one with a nuclear MCA already used in the earlier limit-setting phases of the experiment. We present the performance of the data acquisition system and discuss the relative advantages of the different digitizers.}

\maketitle

\section{Introduction}\label{sec1}

The Beryllium-7 Electron capture in Superconducting Tunnel junctions (``BeEST'') experiment \cite{Leach2022} currently sets leading laboratory-based exclusion limits on heavy neutrinos in the 100-850 keV mass range through precise measurements of $^7$Be decay \cite{Friedrich2021}. For the experiment, $^7$Be is implanted into superconducting tunnel junction (STJ) sensors at the TRIUMF-ISAC rare isotope beam facility, and its decay is measured in an adiabatic demagnetization refrigerator at 100 mK at LLNL. The electron capture decay deposits of order $\sim$100 eV in the STJ from the $^7$Li recoil and the relaxation of its electronic shell \cite{Leach2022}. This produces $\sim$10 nA$_{pp}$ current signals with a rise time of $\sim$1 $\mu$s and a decay time of $\sim$100 $\mu$s that are read out with a specialized transimpedance amplifier from XIA LLC \cite{Warburton2015}. The STJ detector array is calibrated in-situ by illuminating it with a pulsed 355 nm laser (Spectra Physics, model J40-B16-106Q) at a rate of 100 Hz throughout the data acquisition. The number of detected photons varies and produces a comb of spectral lines whose energies are precise multiples of 3.49865 $\pm$ 0.00015 eV \cite{Ponce2018}. In addition, $\gamma$ rays from two of the $^7$Be decay branches can Compton scatter in the Si substrate below the STJ and produce a broad spectral background \cite{Fretwell2020}. Heavy neutrinos would produce an offset $^7$Li recoil spectrum at a reduced energy determined by the neutrino mass with a relative intensity set by its admixture to the electron neutrino. To improve current limits to admixtures $<10^{-5}$, exquisite energy precision and artifact rejection are required.

Phase-III of the BeEST experiment \cite{Leach2022} aims to improve the sensitivity of the heavy neutrino search and evaluate scaling-related challenges for Phase-IV by scaling from a single STJ to a small array of STJ sensors. This introduces the possibility of correlating signals in different pixels and assessing pixel-to-pixel variations. The data acquisition system (DAQ) should provide maximum flexibility to understand spectral details and the needs and trade-offs in future upgrades. Additionally, the digitizer must not contribute to the electronic noise beyond the pre-amplifier contribution of $\sim$1 eV \cite{Marino2022}, should be capable of timing accuracy of order $\sim$1 $\mu$s to tag coincident events, and must demonstrate calibration residuals $<$0.1 eV so that nonlinearity artifacts cannot produce a false heavy neutrino signal in the mass range of interest \cite{Friedrich2021}. This paper discusses the DAQ and its performance in Phase-III.

\section{Data Collection}\label{sec2}

In order to comprehensively analyze the spectral features of the BeEST and assess the full response characteristics of the STJ detectors, we constructed a new DAQ that saves continuous synchronous waveforms from 16 STJ channels. The accumulated data stream is later processed offline, enabling, for the first time, pulse shape analysis of the $^7$Be decay signals. The DAQ is designed around two NI PXIe-6356 cards, each of which digitizes 16-bit differential voltage samples from 8 STJs at a rate of 1.25 MSa/s. The voltage samples are captured on a range of [-1,1] V to match the pre-amplifier output range, and time synchronized with each other by deriving all of the sample clocks from the same 100 MHz backplane clock. Additionally, each PXIe-6356 offers 24 separate digital inputs, one of which is used to synchronously acquire the laser trigger TTL signal. To minimize aliasing, we designed an external differential low pass filter around the Mini-Circuits LPF-B0R35+. In testing, each channel achieved an attenuation of at least 10 dB at 625 kHz. 

\begin{wrapfigure}{r}{0.6001\textwidth}
    \centering
    \vspace*{-0.5cm}
    \includegraphics[width=0.60\textwidth]{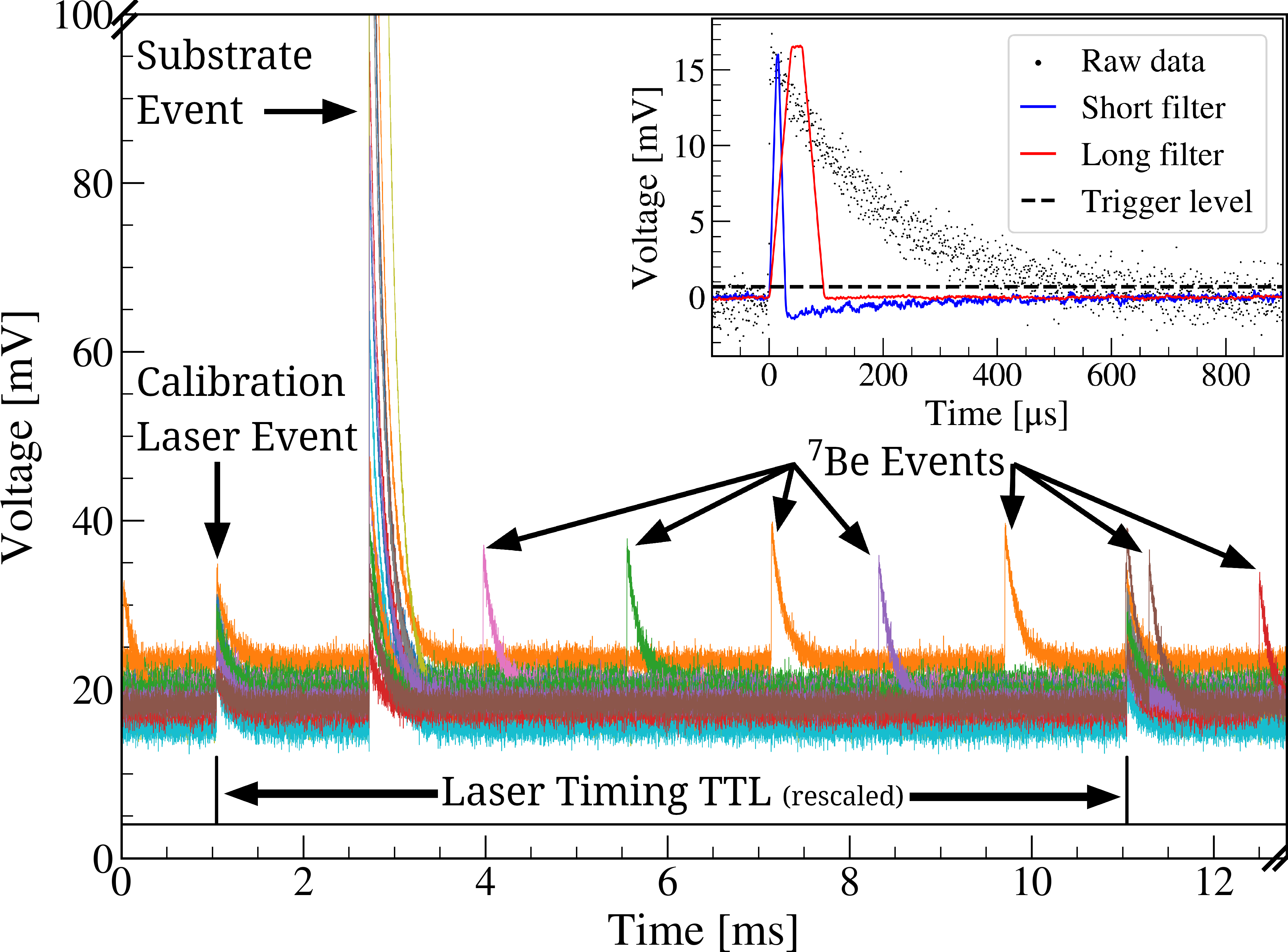}
    \vspace*{-0.3cm}
    \caption{Waveforms from 16 STJ channels showing seven $^7$Be events, two laser pulses, and one substrate event.  Inset: Short and long trapezoidal filters of a sample pulse.\label{fig:traces}} 
    \vspace*{-0.5cm}
\end{wrapfigure}

Over the course of the 50-day acquisition for the Phase-III physics run, the NI DAQ saved 160 TB of data, a sample of which is shown in Figure \ref{fig:traces}. We chose to save waveforms in 10 minute segments for data integrity and ease-of-use, with a few seconds of downtime between segments. While handling the data volume was nontrivial, the data acquisition was particularly robust because there was no on-line processing that could be incorrectly configured.

The waveforms were processed with two trapezoidal filters - a short filter with shaping and flat-top times of 16 and 4 $\mu$s, respectively, and a long filter with 48 and 24 $\mu$s for those values as well as decay time correction. The inset of Figure \ref{fig:traces} shows the results of processing a $^7$Be pulse with these filters. The short filter is used to identify pulses in the data stream by using a rising edge trigger with the trigger level set 5$\sigma$ above the filtered baseline. Next, we fit three line segments to the pre-trigger baseline, rising edge, and flat top of the long filter, and calculate the pulse arrival time as the intersection of the fitted baseline and rising edge. Finally, as an estimate of the event energy, we measure the pulse height by taking the difference between the center of the long filter flat-top and the pre-trigger baseline. 

The four remaining operational STJs are digitized with an XIA MPX-32D \cite{Warburton2015} which is configured to save the trapezoidally filtered pulse maximum and rising-edge trigger time for each pulse in list mode. The trapezoidal filter is calculated by an onboard FPGA, which is fed by four 8-channel 12-bit ADCs that operate at 50 MSa/s. Similar to the continuous DAQ, there are two trapezoidal filter options; however the longer filter is used for both pulse height determination and for detecting pulses in the data stream because its lower noise reduces the minimum trigger level. The longer filter uses shaping and flat-top times of 16 and 1 $\mu$s, and features pixel-specific decay time correction.  In addition to saving data in list mode, the XIA MPX-32D provides the crucial ability to save I(V) curves, digitally control the bias voltage, and analyze real-time MCA spectra from the entire array for setup and diagnostic purposes.

Finally, we duplicate the signal from one STJ at the pre-amplifier output, filter it with an Ortec 672 Shaping Amplifier, and read it out with an Ortec Aspec 927 nuclear MCA. The shaping amplifier is configured with a 10 $\mu$s shaping time and pole-zero correction tuned to match the pulse decay time, and the Ortec Aspec 927 saves two spectra of filtered pulse maxima - one in coincidence with the laser trigger TTL and the other in anticoincidence. A relatively short 90-second MCA window is chosen for flexible drift correction and to reject short periods of increased pickup. Despite being limited to a single channel, the Ortec Aspec 927 is a valuable comparison for the other digitizers because it was the digitizer used in Phase-II of the BeEST experiment and because it features sliding-scale linearization to achieve DAQ nonlinearity as low as 1.6 meV rms \cite{Friedrich2020-nonlinearity,Marino2022}.

\section{DAQ Performance}\label{sec3}

\begin{figure}[ht]
    \centering
    \includegraphics[width=0.94\textwidth]{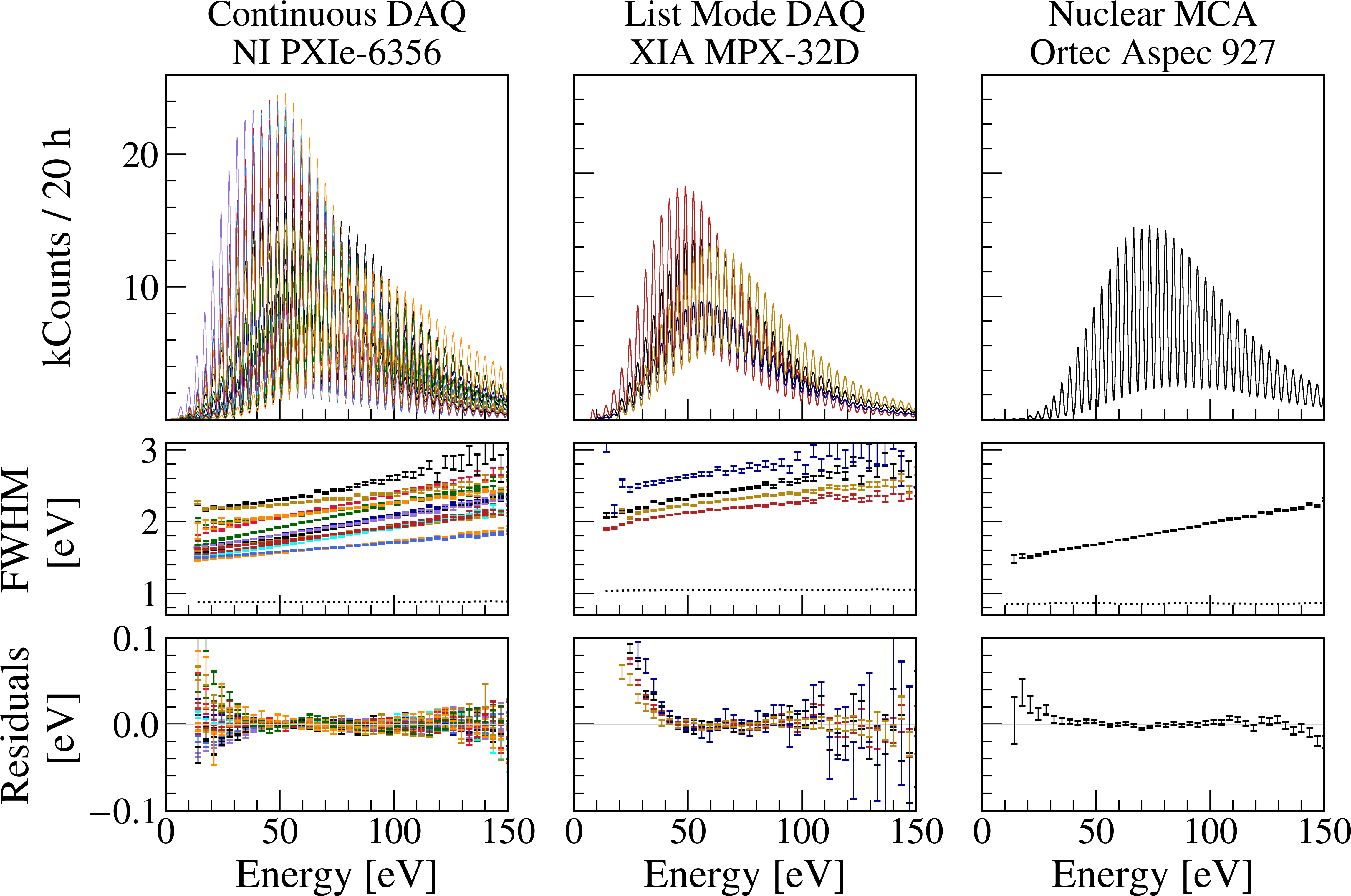}
    \caption{One day of calibrated laser energy spectra (top row), energy resolutions (middle row), and calibration residuals (bottom row) from the NI PXIe-6356 (left, 16 channels in different colors), XIA MPX-32D (middle, 4 channels) and Ortec Aspec 927 (right). The dotted lines show AWG results.\label{fig:nonlinearity}}
    \vspace*{-1em}
\end{figure}

To calibrate the data, we use \texttt{iminuit} \cite{iminuit} to simultaneously fit several Gaussian functions to the laser spectrum from each 10 minute segment, with the their centroids providing energy calibration points. A second-order polynomial is fit to the centroids in the energy range from 42 to 147 eV and used to calibrate that segment. The calibrated segments are summed and fit once again with a comb of Gaussian functions to more precisely extract the centroid and width of each laser peak, which are respectively used to measure the energy linearity and resolution of the system. Specifically, the FWHM of the Gaussians measures the energy resolution of each STJ pixel as a function of energy and the Gaussian centroids are used to determine the fit residuals from a second order polynomial, both of which are shown in Figure \ref{fig:nonlinearity}. In the calibration range, the NI PXIe-6356 has a mean residual of 5.9 meV rms with a channel-to-channel variation of $\pm$ 2.5 meV. The MPX-32D has residuals of 9.5 $\pm$ 4.0 meV rms, and the Ortec Aspec 927 has a calibration residual of only 3.4 meV rms on its single channel. The residuals of the Ortec DAQ are higher than the 1.6 meV rms it has previously demonstrated \cite{Friedrich2020-nonlinearity} due to reduced statistics and uncertainty in detector drift correction.

\begin{figure}[ht]
\begin{minipage}[c]{0.495\linewidth}
  \includegraphics[width=\textwidth]{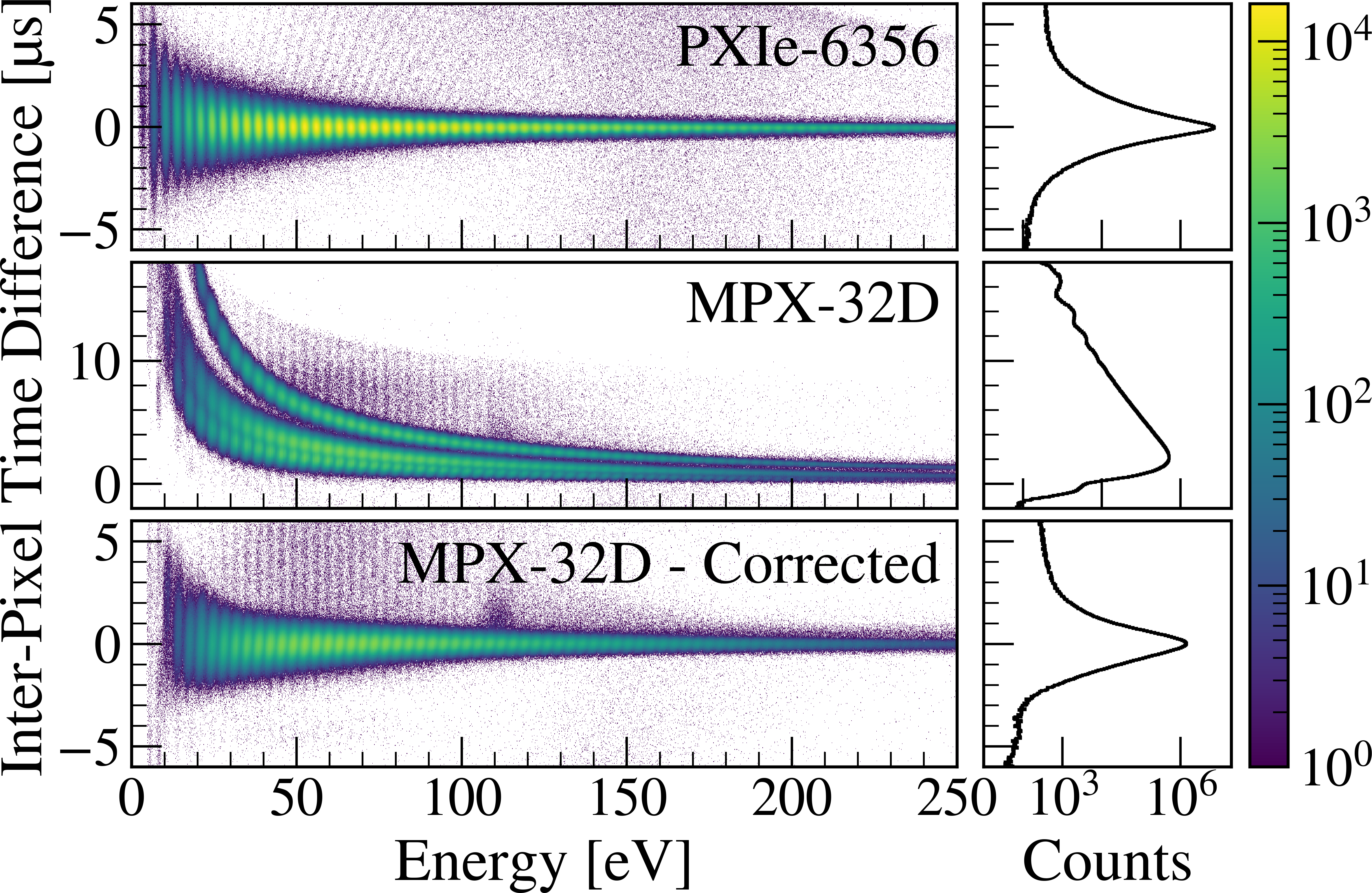}
  \vspace*{-1em}
  \caption{Time distributions between simultaneous laser events from the NI (top) and XIA (middle) DAQs in log scale. Channels with different trigger levels create multiple bands in the middle plot, which are corrected for in the bottom plot. Events outside the central bands are $^7$Be events in random coincidence with laser events.\label{fig:timingHeatmap}}
\end{minipage}
\hfill
\begin{minipage}[c]{0.495\linewidth}
  \includegraphics[width=\textwidth]{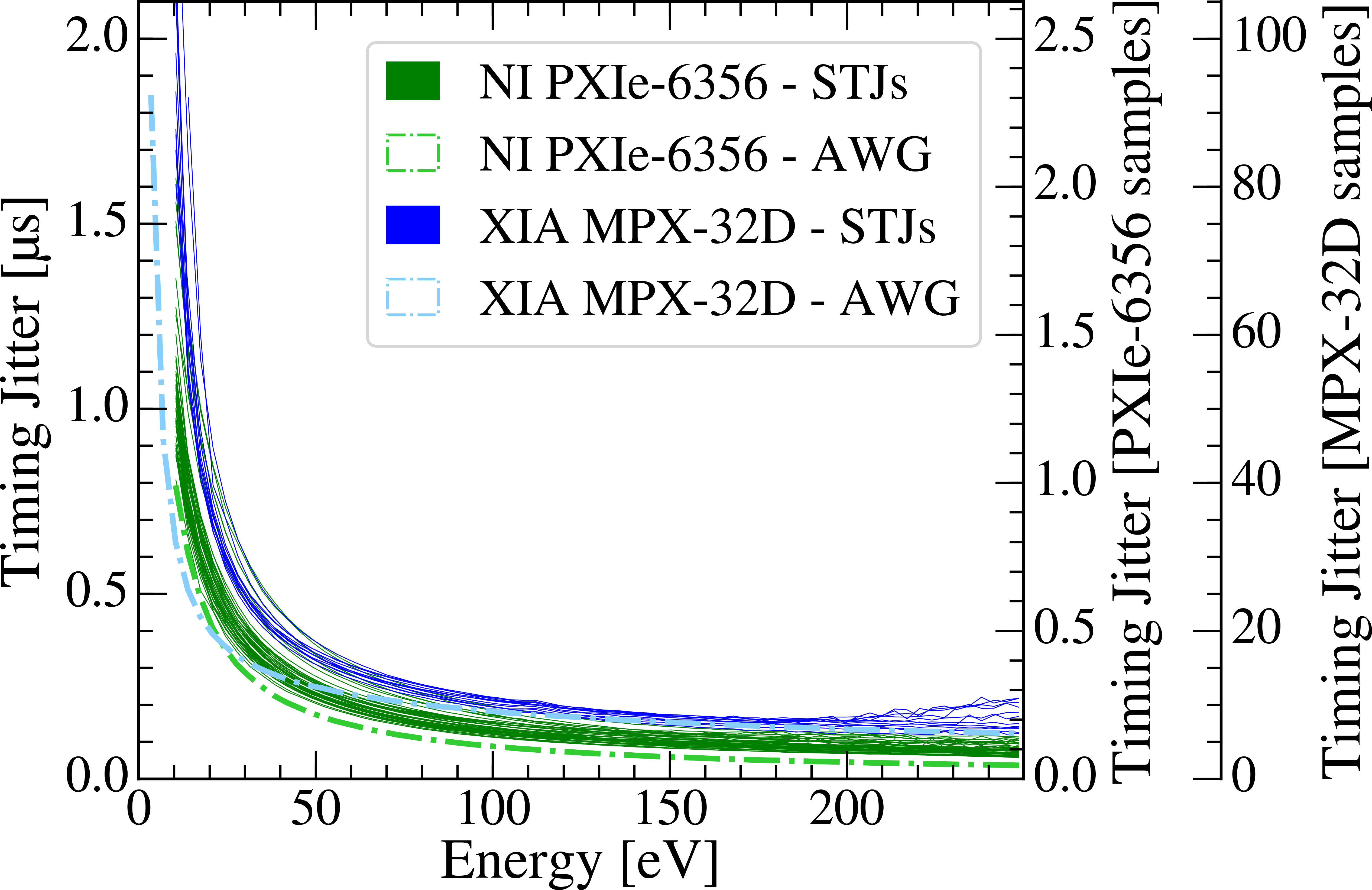}
  \vspace*{-1em}
  \caption{Comparison of the timing jitter of the NI and XIA DAQs across the three days of unblinded data (solid lines represent different channels). The increase at high energies for some channels is due to low statistics and increased pile-up fraction. Comparison with AWG pulses is shown for both DAQs (dashed lines), assuming 150 mV/keV.\label{fig:timingJitterPlot}}
\end{minipage}%
\end{figure}

The timing accuracy of the NI and XIA DAQs is evaluated by calculating the measured arrival time differences between nominally simultaneous laser signals in different pixels, which are shown in Figure \ref{fig:timingHeatmap}. As expected for a rising edge trigger \cite{Knoll}, the MPX-32D displays energy walk due to the finite signal rise time. The energy walk is fit to and corrected using \texttt{iminuit} \cite{iminuit}. Afterwards, pile-up and random coincidence events are removed by rejecting events with a time difference $>$3$\sigma$, and the standard deviation of the remaining events from each laser peak is plotted in Figure \ref{fig:timingJitterPlot}. We observe that the NI PXIe-6356 can achieve a timing jitter much smaller than its sampling rate, which is possible because the pulse arrival time is determined using data from the entire pulse, rather than just a single point on the rising edge. Its timing jitter is also slightly better than the MPX-32D, despite a 40 times slower sampling rate. That is in part because we chose to trigger using the long filter on the MPX-32D in order to achieve a lower energy threshold at the expense of a degraded timing resolution.


Finally, an NI PXIe-5423 arbitrary waveform generator (AWG) is connected to the pre-amplifier input to evaluate the intrinsic energy and timing resolution of each DAQ using a comb of artificial STJ-like pulses. The same methods as above were used to extract the energy resolution and timing jitter, which are compared with the real STJ signals in Figures \ref{fig:nonlinearity} and \ref{fig:timingJitterPlot}.

\section{Discussion and Conclusions}\label{sec4}

To determine the performance of each DAQ, we evaluated the resolution, calibration residuals, and timing jitter of each channel using the small set of unblinded data and AWG tests. First, the DAQ resolution is measured to be $<$ 1.1 eV for all three digitizers, which meets our requirements and indicates that the channel-to-channel resolution variations are due to STJ device differences (Figure \ref{fig:nonlinearity}). Second, both the XIA MPX-32D and the NI PXIe-6356 meet our timing resolution target of $\sim$1 $\mu$s, with slightly higher timing jitter only near the trigger threshold outside the energy range of primary interest (Figure \ref{fig:timingHeatmap}, \ref{fig:timingJitterPlot}). This is particularly impressive for the NI PXIe-6356, which is able to achieve sub-sample timing resolution because of the pulse arrival time extraction algorithm. On the other hand, the Ortec Aspec 927 does not meet the timing or channel number needs for Phase-III, which limits the types of analysis that can be performed with its data. Nonetheless, it is still a valuable comparison to the Phase-II DAQ. Third, the calibration residuals are typically $\lesssim$10 meV rms in the calibration range and thus well below the 0.1 eV target. In summary, we have demonstrated that two of the three DAQ setups meet the requirements for Phase-III of the BeEST experiment, and expect that the ultimate sensitivity for the physics goals will be limited by other systematic uncertainties. 

Still, improvements to the DAQ are highly desirable as we scale up to $>$100 channels for Phase-IV \cite{Leach2022}. First, combining the pulse arrival time algorithm with a faster sampling rate would significantly improve the timing resolution. That would allow for shorter coincidence windows and improved pile-up identification, which is critical at increased count rates. Additionally, capturing waveforms with a faster ADC could enable rise-time analysis to discriminate between top and bottom electrode events \cite{NAKAMURA200011}. To mitigate the increase in data volume from increasing both the sampling rate and number of channels, an ideal Phase-IV DAQ would save only triggered waveforms. The data throughput could be further reduced by recording the pre- and post-trigger sections of each pulse at a slower sampling rate compared to the rising edge. Furthermore, as we scale to a larger number of detectors, manual per-device per-cooldown processes need to be minimized. These include optimizing the STJ bias point, STJ zero, decay time correction, and trigger level. Finally, real-time monitoring will be necessary to quickly identify poorly performing devices, and the DAQ control software needs to be robust and error tolerant.

We have collected a wealth of data from the three DAQs and are exploring new analysis techniques enabled by the multi-channel setup and the continuous DAQ. This includes pulse shape analysis to extract an experimental pile-up spectrum, substrate event tagging, and trigger efficiency studies. Using those techniques and the small set of currently unblinded data, we are in the process of developing robust analysis tools to project Phase-III's sensitivity to sub-MeV neutrinos.

\backmatter

\bmhead{Acknowledgments}
C. Bray gratefully acknowledges support from the DOE Office of Science SCGSR program. The BeEST experiment is funded by the Gordon and Betty Moore Foundation (10.37807/GBMF11571), the DOE-SC Office of Nuclear Physics (DE-SC0021245, DE-FG02-93ER40789, and SCW1758), and the LLNL LDRD grant 20-LW-006. F. Ponce is supported by PNNL under contract DE-AC05-76RL01830. TRIUMF receives federal funding via a contribution agreement with the National Research Council of Canada. The theoretical work was performed as part of the European EMPIR Projects 17FUN02 MetroMMC and 20FUN09 PrimA-LTD. This work was performed under the auspices of the U.S. Department of Energy by LLNL under Contract No. DE-AC52-07NA27344.

\bibliography{sn-bibliography}

\end{document}